\documentclass[12pt]{article}

\usepackage{graphicx}

\def\um{{1}/{2}}
\def\sq{1/\sqrt{2}}

\begin{document}
\renewcommand{\thefootnote}{\fnsymbol{footnote}}
\renewcommand{\theenumi}{(\roman{enumi})}
\title{The neutrinoless double $\beta$ decay \\ 
 and the neutrino mass hierarchy}
\author{
{Naoyuki Haba$^{1,2}$}\thanks{haba@eken.phys.nagoya-u.ac.jp}
{, and Tomoharu Suzuki$^2$}\thanks{tomoharu@eken.phys.nagoya-u.ac.jp}
\\
\\
\\
{\small \it $^1$Faculty of Engineering, Mie University,}
{\small \it Tsu Mie 514-8507, Japan}\\
{\small \it $^2$Department of Physics, Nagoya University,}
{\small \it Nagoya, 464-8602, Japan}\\
}
\date{}
\maketitle

\vspace{-9.5cm}
\begin{flushright}
hep-ph/0202143\\
DPNU-02-03\\
\end{flushright}
\vspace{9.5cm}
\vspace{-2.5cm}
\begin{center}
\end{center}
\renewcommand{\thefootnote}{\fnsymbol{footnote}}

\begin{abstract}

Recently 
 the evidence of the neutrinoless double $\beta$ 
 ($0\nu \beta\beta$) decay has been announced. 
This means that neutrinos are Majorana particles and 
 their mass hierarchy is forced 
 to certain patterns in the diagonal basis of
 charged lepton mass matrix. 
We estimate the magnitude of $0\nu \beta\beta$ decay 
 in the classification of the neutrino mass hierarchy patterns 
 as Type A, $m_{1,2} \ll m_{3}$, Type B, 
 $m_1 \sim m_2 \gg m_3$, 
 and Type C, $m_1 \sim m_2 \sim m_3$,
 where $m_{i}$ is the $i$-th generation neutrino absolute mass. 
The data of $0\nu \beta\beta$ decay experiment 
 suggests the neutrino mass hierarchy pattern should be 
 Type B or C. 
Type B predicts a small magnitude of 
 $0\nu \beta\beta$ decay which is 
 just edge of the allowed region of 
 experimental value in $95\%\;\mathrm{c.l.}$,  
 where Majorana $CP$ phases 
 should be in a certain parameter region. 
Type C can induce the suitably large amount 
 of $0\nu \beta\beta$ decay which is consistent 
 with the experimental data, 
 where overall scale of degenerate neutrino mass 
 plays a crucial role, and its large value 
 can induce the large $0\nu \beta\beta$ decay
 in any parameter regions of 
 Majorana $CP$ phases.

\end{abstract}

\newpage
\section{Introduction}

Recently the evidences of neutrino oscillations are strongly supported
 by both of the atmospheric~\cite{Kamiokande,SKatm} 
 and the solar neutrino experiments~
\cite{SKsolar,SNO,Cl-Homestake,Ga-Gallex-and-GNO}. 
The former suggests an almost maximal lepton flavor mixing 
 between the 2nd and the 3rd generations, 
 while the favorable solution to the solar neutrino deficits 
 is given by large mixing angle solution between the
 1st and the 2nd generations ( LMA, LOW or VO )~\cite{post-SNO-analysis}.
Neutrino oscillation experiments indicate that the neutrinos have tiny
 but finite masses, with two mass squared differences 
 $\Delta m_{\odot}^2< \Delta m_{\rm atm}^2$.  
However, 
 we cannot know 
 the absolute values of the neutrinos masses 
 from the oscillation experiments. 

Recently, 
 a paper\cite{db} 
 announces the evidence of 
 the neutrinoless double $\beta$ 
 ($0\nu \beta\beta$) decay.  
This paper suggests 
\begin{eqnarray}
\langle m \rangle=(0.05-0.86) \mathrm{eV}\quad at\;95\%\;\mathrm{c.l.}
\quad\mathrm{(best\;value\;0.4\;eV)}\;.
\label{result}
\end{eqnarray}
This results is very exciting. 
It is because 
 $0\nu\beta\beta$ decay experiments
 could tell us about the absolute value of the neutrino masses, 
 while neutrino oscillation experiments 
 show only mass squared differences of neutrinos. 
The evidence for $0\nu\beta\beta$ decay 
 also means neutrinos are the Majorana particles and 
 the lepton number is violated, since 
 $0\nu\beta\beta$ decay cannot occur in the case of 
 Dirac neutrinos. 
This evidence is also closely related to 
 the recent topics of the cosmology such as 
 the dark matter candidate of universe\cite{dark}. 

The tiny neutrino masses and 
 the lepton flavor mixings have 
 been discussed in 
 a lot of models beyond the Standard Model (SM). 
One of the most promising ideas is 
 that light neutrinos are constructed as Majorana particles 
 in the low energy, such as 
 the see-saw mechanism\cite{seesaw}. 
Here we are concentrating on 
 the light Majorana neutrinos 
 which masses are induced by the 
 dimension five operators in the 
 low energy effective Yukawa interactions. 

In this paper, 
 we will estimate the magnitude of $0\nu \beta\beta$ decay
 in the classification of the neutrino mass hierarchy patterns 
 as Type A, $m_{1,2} \ll m_{3}$, Type B, 
 $m_1 \sim m_2 \gg m_3$, 
 and Type C, $m_1 \sim m_2 \sim m_3$,
 where $m_{i}$ is the $i$-th generation neutrino absolute mass
\cite{Altarelli}. 
The magnitude of $0\nu \beta\beta$ decay 
 strongly depends on the neutrino mass hierarchy. 
We will analyze the value of $\langle m \rangle$ 
 in the parameter space of the Majorana 
 $CP$ phases, and search the region of 
 being consistent with 
 the data of $0\nu \beta\beta$ decay experiment.



%
\section{Neutrino masses and flavor mixings}

Before starting our discussions, 
 let us show the notations of neutrino 
 masses and their mixing angles. 
Since we would like to be 
 concentrating on the low energy mass matrix 
 of Majorana neutrinos,  
 we add the dimension five operator
 to the standard model in the lepton sector as,   
\begin{eqnarray}
\mathcal{L}=(Y_e)_{ij}L_iE_jH_1+
\kappa_{ij}(L_i H_2)(L_j H_2)\quad,
\label{dim5}
\end{eqnarray}
where $\kappa_{ij}$ is the coefficient of 
 the dimension five operator, and  
 $L_i$ and $E_i$ are $i$-th generation
lepton doublet and charged lepton singlet with 
 $(i=1 \sim 3)$,
 respectively,
$H_1$ and $H_2$ are Higgs doublets
\footnote{
Although we write Yukawa interactions in Eq.(\ref{dim5}) 
 by the supersymmetric forms, 
 the results in this paper 
 do not depend on whether the model is supersymmetric or not.
}.
The light neutrinos obtain masses
 from the vacuum expectation value (VEV) of 
 $v_2=\langle H_2 \rangle$ as 
\begin{eqnarray}
(M_\nu)_{ij}
=\kappa_{ij}v_2^2\quad,
\end{eqnarray}
The energy scale of $\kappa_{ij}^{-1}$ 
 suggests the scale of the new physics and 
 the lepton number 
 violation. 
In the diagonal base of the charged lepton sector, 
 the light neutrino mass matrix, $(M_\nu)_{ij}$, 
 is diagonalized by $U_{ij}$
 as
\begin{eqnarray}
U^{\mathrm{T}}M_\nu U \equiv M_\nu^{diag}
=diag(m_1,m_2,m_3)\quad.
\label{MM}
\end{eqnarray}
Here the matrix $U$ is so-called MNS matrix\cite{MNS} 
 denoted by 
\begin{eqnarray}
U=V \cdot P\quad,
\label{MNSP}
\end{eqnarray}
where $V$ is the CKM-\textit{like} matrix,
which contains one $CP$-phase ($\delta$),
\begin{eqnarray*}
V=\left(
\begin{array}{ccc}
c_{13}c_{12}&c_{13}s_{12}&s_{13} e^{-i\delta}\\
-c_{23}s_{12}-s_{23}s_{13}c_{12}e^{i\delta}&
c_{23}c_{12}-s_{23}s_{13}s_{12}e^{i\delta}&
s_{23}c_{13}\\
s_{23}s_{12}-c_{23}s_{13}c_{12}e^{i\delta}&
-s_{23}c_{12}-c_{23}s_{13}s_{12}e^{i\delta}&
c_{23}c_{13}
\end{array}
\right)
\end{eqnarray*}
and 
$P$ contains two extra Majorana phases ($\phi_{1,2}$),
\begin{eqnarray*}
P=diag.(e^{-i {\phi_1/2}},e^{-i \phi_2/2},1)\quad.
\end{eqnarray*}
For the latter convenience, 
we introduce the matrix 
\begin{eqnarray}
\widetilde{M}_\nu\equiv 
P^* M_\nu^{diag} P^{*}=diag.(m_1e^{i \phi_1},m_2 e^{i\phi_2},m_3)
=diag.(\tilde{m}_1,\tilde{m}_2,\tilde{m}_3)\quad.
\label{tildeM}
\end{eqnarray}

Table \ref{table} shows the results from 
 the recent neutrino oscillation experiments
\cite{
Kamiokande,SKatm,SKsolar,SNO,Cl-Homestake,
Ga-Gallex-and-GNO,post-SNO-analysis}. 
\begin{table}
\begin{center}
\begin{tabular}{||c||cc||}\hline\hline
\textit{Solar}&$\Delta m_{\odot}^2(\mathrm{eV^2})$&$\sin^2 2\theta_{\odot}$
\\\hline
SMA&$(4-9)\times 10^{-6}$&$(0.0008-0.008)$\\
LMA&$(2-20)\times 10^{-5}$&$(0.3-0.93)$\\
LOW&$(6-20)\times 10^{-8}$&$(0.89-1)$\\
VO&$10^{-10}$&$(0.7-0.95)$\\
Just So&$(5-8)\times 10^{-12}$&$(0.89-1)$\\\hline\hline
\textit{Atmospheric}&$\Delta m_{atm}^2(\mathrm{eV^2})$
&$\sin^2 2\theta_{{atm}}$
\\\hline
&$(1.8-4.0)\times 10^{-3}$&$(0.87-1)$\\\hline\hline
\end{tabular}
\end{center}
\caption{
The allowed values of 
$\Delta m_{\odot}^2$, $\Delta m_{atm}^2$, $\sin^22\theta_{\odot}$, 
and $\sin^22\theta_{{atm}}$
from the neutrino oscillation experiments
\cite{Kamiokande,SKatm,SKsolar,SNO,Cl-Homestake,
Ga-Gallex-and-GNO,post-SNO-analysis}.
}\label{table}
\end{table}
These results 
 indicate that the neutrinos have tiny
 but finite masses, with two mass squared differences 
 $\Delta m_{\odot}^2< \Delta m_{ atm}^2$.  
The naive explanation of 
 the present 
 neutrino oscillation 
 experiments is that 
 the solar neutrino anomaly is caused by 
 the mixings of 
 the 1st and the 2nd generations 
 ($\theta_{\odot}\simeq\theta_{12}$, 
 $\Delta m^2_{\odot}\simeq m_{2}^2-m_1^2$), 
 and atmospheric neutrino deficit is caused by  
 the mixings of the 2nd and the 3rd generations
 ($\theta_{atm}\simeq\theta_{23}$, 
 $\Delta m^2_{atm}\simeq m_3^2-m_2^2$).
Considering the results of the oscillation experiments,
 the hierarchical patterns of neutrino masses 
 are classified by the following three types: 
\begin{center}
\begin{tabular}{ccl}
A&:&$m_3\gg m_{1,2}$\\
B&:&$m_1\sim m_2\gg m_3$\\
C&:&$m_1 \sim m_2 \sim m_3$ \quad .
\end{tabular}
\end{center}
Taking into account of the mass squared differences, 
 $\Delta m_{\odot}^2$ and $\Delta m_{atm}^2$,
 the absolute 
 masses of the neutrino in the 
 leading are written by 
\begin{eqnarray}
\begin{minipage}{13.5cm}
\begin{flushleft}
\begin{tabular}{llcl}
\underline{Type A}&\\
&$m_1$&:&$0$\\
&$m_2$&:&$\sqrt{\Delta m_{\odot}^2}$\\
&$m_3$&:&$\sqrt{\Delta m_{atm}^2}$
\end{tabular}
\begin{tabular}{llcl}
\underline{Type B}&\\
&$m_1$&:&$\sqrt{\Delta m_{atm}^2}$\\
&$m_2$&:&$\sqrt{\Delta m_{atm}^2}+\frac{1}{2}\frac{\Delta m_{\odot}^2}{\sqrt{\Delta m_{atm}^2}}
$\\
&$m_3$&:&$0$
\end{tabular}
\begin{tabular}{llcl}
\underline{Type C}&\\
&$m_1$&:&$m_\nu$\\
&$m_2$&:&$m_\nu+\frac{1}{2}\frac{\Delta m_{\odot}^2}{m_\nu}$\\
&$m_3$&:&$m_\nu
+\frac{1}{2}\frac{\Delta m_{atm}^2}{m_\nu}$\quad ,
\end{tabular}
\end{flushleft}
\end{minipage}
\label{masstype}
\end{eqnarray}
in each type, respectively. 
Where $m_\nu$ in Type C is the scale of 
 the degenerated neutrino masses.

Table 1 shows 
 the flavor mixing angles in the lepton sector.   
The mixing angle between the 2nd and the 
 3rd generations is almost maximal mixing 
 from the atmospheric neutrino experiments. 
There are three candidates
\footnote{
Since there are no defferences between
the three solutions LOW, VO and Just So, 
we consider them together VAC solution.
}
 of solutions 
 for solar neutrino problems. 
Recent super-kamiokande data of day/night effects implies 
 the LMA solution are most favorable solution
\cite{SKsolar}. 
We use the center values of the 
 mixing angles $\theta_{12}$ 
 and $\theta_{23}$, 
\begin{center}
\begin{minipage}{5cm}
\begin{eqnarray*}
&&\Delta m_{atm}^2=3.2\times 10^{-3}\;\mathrm{eV^2}\;,\\
&&\sin^2 2\theta_{atm}=1.0\;,
\end{eqnarray*}
\end{minipage}
\hspace{1cm}
\begin{minipage}{5cm}
\begin{eqnarray*}
&&\Delta m_{\odot}^2=4.5\times 10^{-5}\;\mathrm{eV^2}\;,\\
&&\tan^2\theta_{\odot}=4.1\times 10^{-1}\;,
\end{eqnarray*}
\end{minipage}
\end{center}
in the following numerical analyses. 
The rest of the mixing angle, $\theta_{13}$, 
 is not measured as the deficit value, 
 the upper bound is given as, 
\begin{eqnarray*}
\sin^2 2\theta_{13}<0.1,
\end{eqnarray*}
{} from the CHOOZ experiments\cite{CHOOZ}.

In the zeroth order approximations, 
 $\displaystyle \cos \theta_{12}=\cos\theta_{23}={1}/{\sqrt{2}}$ and 
 $\sin\theta_{13}=0$, 
 we can obtain 
 the zeroth order form of the 
 MNS matrix\footnote{
The deviation from this zeroth order approximation is important. 
As will be shown in the next section, 
 the tiny difference between the maximal and the LMA angle 
 causes the sensitive effects to 
 the mass neutrino matrix. 
} as, 
%
%
\begin{eqnarray}
V^{(0)}=\left(
\begin{array}{ccc}
\frac{1}{\sqrt{2}}&\frac{1}{\sqrt{2}}&0\\
-\frac{1}{{2}}&\frac{1}{{2}}&\frac{1}{\sqrt{2}}\\
\frac{1}{2}&-\frac{1}{2}&\frac{1}{\sqrt{2}}
\end{array}
\right). 
\label{Vzero}
\end{eqnarray}
The neutrino mass matrix $M_\nu$ is 
 determined by ${}U$ and ${M}_{\nu}^{diag}$ 
 from Eqs.(\ref{MM})$\sim$(\ref{tildeM}). 
The zeroth order form of the neutrino mass 
 matrix is determined by the approximated MNS matrix, $V$, 
 according to the patterns of neutrino mass hierarchy,
 Types A$\sim$C.    
In Ref.\cite{Altarelli}, 
 the zeroth order forms of neutrino mass matrices 
 are shown when Majorana $CP$ phases are $0$ or 
 $\pi$, which are shown in Table 2. 
These mass matrices are 
 useful for the first approximations 
 of estimating the probability of $0\nu\beta\beta$. 
We use continuous values of Majorana $CP$ phases 
 in the following analyses.



\section{Neutrino mass hierarchies and $0\nu\beta\beta$ decay}

Now we are in a position of discussing 
 the relations between the neutrino mass hierarchy 
 patterns and the magnitude of $0\nu\beta\beta$ decay. 
The effective neutrino mass $\langle m\rangle$, which shows 
 the magnitude of $0\nu\beta\beta$ decay
 in Eq.(\ref{result}) 
 is defined by
\begin{eqnarray}
\langle m \rangle&=& |\sum_{i=1}^{3}U_{ei}^2 m_i |
=|\sum_{i=1}^{3}U_{ei} m_i U^{\mathrm{T}}_{ie} |
=|V_{e1}^2 m_1 e^{i\phi_1}+V_{e2}^2 m_2 e^{i\phi_2}+V_{e3}^2 m_3|,
\label{<m>}
\end{eqnarray}
where $i$ denotes the label of the mass 
 eigenstate ($i=1,2,3$). 
What pattern of neutrino mass hierarchy  
 can induce the sizeable amount of $0\nu\beta\beta$ decay
 amplitude in Eq.(\ref{result})? 
In the approximation in the previous section, 
 the value of $\langle m \rangle$ 
 is equal to that of $(1,1)$ component of $\widetilde{M}_\nu$. 
Thus, Table 2 suggests that 
 the forms of the neutrino mass matrix should be 
 B2 or C0 or C3,  
 in order to obtain the suitable 
 large magnitude 
 of $(1,1)$ component. 
However, it is too naive estimation. 
We will estimate the magnitude of $\langle m\rangle$
 more accurately, 
 which patterns of neutrino mass hierarchies 
 are really consistent with $0\nu\beta\beta$ decay 
 data in Eq.(\ref{result}) within the fluctuations 
 of physical parameters of a tiny value of $|U_{e3}|(=\sin \theta_{13})$, 
 CKM-\textit{like} phase $\delta$ and 
 Majorana phases $\phi_{1,2}$.
We will show the analyses according to 
 the neutrino mass hierarchy patterns, Type A$\sim$C in 
 the base of Majorana $CP$ phases.

\subsection{Type A}

Equation (\ref{<m>}) suggests 
 the largest value of $\langle m \rangle$ is given by ,
\begin{eqnarray}
\langle m \rangle_{\mathrm{MAX}}
=|V_{e1}|^2m_1+|V_{e2}|^2m_2+|V_{e3}|^2m_3\;.
\label{max}
\end{eqnarray}
In the Type A mass hierarchy, 
 the main contribution to $\langle m \rangle$ is 
 2nd term or 3rd term of Eq.(\ref{max}) 
 depending on the magnitude of $V_{e3}$. 
However, both terms induce $\mathcal{O}(10^{-3})\;\mathrm{eV}$
 contribution to the value of $\langle m \rangle$ 
 within the fluctuations of physical 
 parameters, $V_{e3}$. 
Even if $m_1$ in Eq.(\ref{max}) is lifted as 
 $m_1 \sim \sqrt{\Delta m_{\odot}^2} \leq m_2$, 
 the magnitude of $\langle m \rangle$ is still less than 
 of $\mathcal{O}(10^{-2})$ eV. 
These magnitudes are 
 too small to explain the recent $0\nu\beta\beta$ decay
 experiments in Eq.(\ref{result}).
Thus we can conclude that 
 the Type A neutrino mass hierarchy pattern can 
 not explain the $0\nu\beta\beta$ 
 results in Eq.(\ref{result}).

 \subsection{Type B}

Next, 
 let us see the case of Type B.
When Majorana $CP$ phases take the specific values 
 of $0$ or $\pi$, 
 Type B is devided into two cases B1 and B2, 
 which are shown in Table 2. 
\begin{eqnarray*}
(\tilde{m}_1,\tilde{m}_2,\tilde{m}_3)=\left\{
\begin{array}{lrccccl}
\sqrt{\Delta m_{atm}^2}
(1,&-1,&0)&\quad&(\phi_1,\phi_2)=(0,\pi)&\quad&(\mathrm{Type B1})\\
\sqrt{\Delta m_{atm}^2}
(1,& 1,&0)&&(\phi_1,\phi_2)=(0,0)&&(\mathrm{Type B2})
\end{array}
\right.
\end{eqnarray*}
Where mass eigenvalues show the zeroth order 
 values. 
The sign of $m_1$ is the opposite (same) as that of 
 $m_2$ in Type B1 (B2). 
Table 2 shows that B2 induces 
$\langle m \rangle=\mathcal{O}(\sqrt{\Delta m_{atm}^2})$,
while 
 that of B1 does not in the zeroth order. 
Thus we can expect the sizeable amplitude of 
  $0\nu\beta\beta$ decay in B2, not B1. 
Since the Majorana phases, $\phi_{1,2}$, connect 
 B1 and B2 continuously,  
 we estimate the value of 
 $\langle m \rangle$ with nontrivial Majorana phases, 
 and see also the effects of other small quantities 
 such as $U_{e3}$.  

\begin{figure}
\setlength{\unitlength}{1cm}
\begin{picture}(13,7)
\put(0,1){\includegraphics[scale=.7]{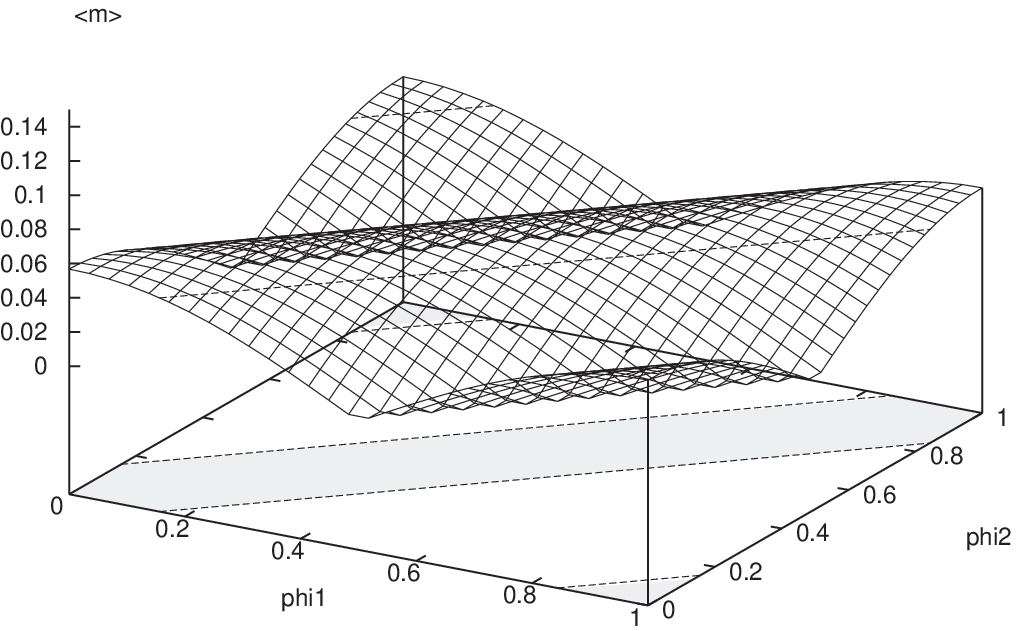}}
\put(4,0){(a)}
\put(8,1){\includegraphics[scale=.7]{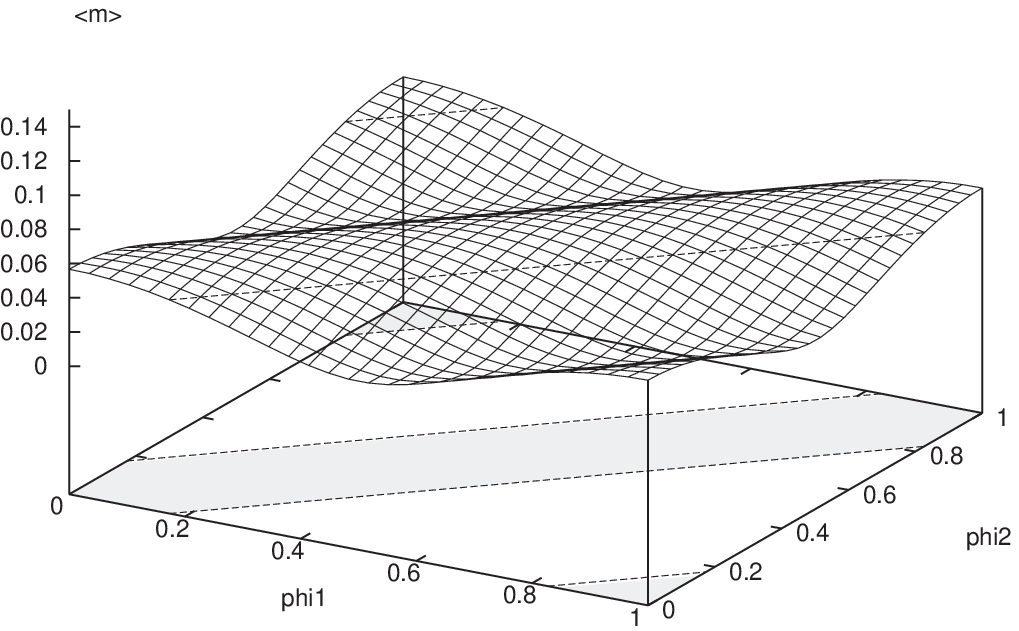}}
\put(12,0){(b)}
\end{picture}
\caption{
The values of $\langle m \rangle$ 
 with the 1-2 maximal mixing, $\tan \theta_{\odot}=0$, in 
 (a), and the LMA center value, $\tan \theta_{\odot}=2.6 \times 
 10^{-1}$, in (b). 
Other values are taken as the zeroth order values as 
 $U_{e3}=0$ and $\theta_{atm}=\pi/4$. 
Majorana phases $\phi_{1,2}$ are normalized by $2\pi$, and  
 the lines $\phi_1-\phi_2=n+1/2$ ($\phi_1-\phi_2=n$)
 correspond to Type B1 (B2) 
 where $n=0,\pm 1,\pm 2\cdots$. 
The region of the 
 large value of $\langle m \rangle$ satisfying 
 Eq.(\ref{result}) is shown 
 as the grayed drown region. 
}
\label{Type B1}
\end{figure}

At first, we estimate 
 the effect of the tiny difference 
 of LMA mixing angle $\theta_{\odot}$ 
 from 
 the 
 maximal mixing between the 1st and the 2nd 
 generations. 
The recent oscillation experiments 
 show the LMA solution disfavor the complete maximal mixing. 
Figure \ref{Type B1} shows 
 the values of $\langle m \rangle$ 
 with the 1-2 maximal mixing, $\tan \theta_{\odot}=0$ in 
 (a), and the LMA center value, $\tan \theta_{\odot}=2.6 \times 
 10^{-1}$, in (b). 
Other values are taken as 
 $U_{e3}=0$ and $\theta_{atm}=\pi/4$. 
The lines where $\phi_1-\phi_2=(2n+1)\pi$ ($\phi_1-\phi_2=2n\pi$)
 correspond to Type B1 (B2) 
 where $n=0,\pm 1,\pm 2\cdots$. 
Only one degree of freedom, $\phi_1-\phi_2$, is 
 physical since 
 $m_3 =0$ in Type B (see Eq.(\ref{tildeM})).
The region of the 
 large value of $\langle m \rangle$ satisfying 
 Eq.(\ref{result}) is shown 
 as the grayed drown region. 
Apparently only the region around B2 
 can satisfy Eq.(\ref{result}). 
The wide region include B1 does not
 satisfy Eq.(\ref{result}). 
Figure \ref{Type B1} shows that Majorana phases connect B1 and B2  
 continuously. 
Comparing to (b) with (a) in Fig.\ref{Type B1}, 
 we find that the small change from the maximal mixing,
 makes an allowed region,   
 where the value $\langle m \rangle$ is 
 larger than $0.05\;\mathrm{eV}$,  
 be spreaded. 
Especially for the values of $\langle m \rangle $, 
 although they 
 are almost zero in the region around 
 Type B2 
 in (a), 
 they become significantly enhanced 
 in the same region in (b). 
This means that 
 the tiny deviation of $\theta_{\odot}$ 
 from the zeroth order classification of neutrino mass matrices, 
 in Table \ref{table:mass}, is important. 
It is because 
 the accidental cancellation 
 between $U_{e1}$ and $U_{e2}$ induced from 
 two maximal mixings, $\theta_{\odot}=\theta_{atm}=\pi/4$, in (a) 
 is forbidden in (b).   
In the following analyses 
 we use the $\theta_{\odot}=2.6 \times 
 10^{-1}$ as LMA angle, 
 not maximal mixing.

\begin{figure}
\setlength{\unitlength}{1cm}
\begin{picture}(13,6.5)
\put(0,1){\includegraphics[scale=.7]{Figure1b.eps}}
\put(4,0){(a)}
\put(8,1){\includegraphics[scale=.7]{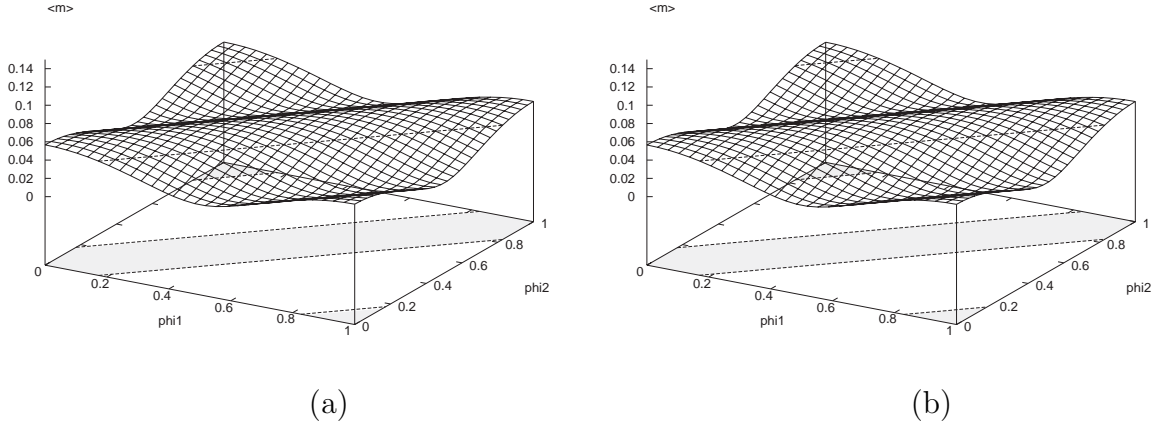}}
\put(12,0){(b)}
\end{picture}
\caption{
The values of $\langle m \rangle$ 
 with $|U_{e3}|=0$ (a) 
 and $|U_{e3}|=0.1$ (b).
Other parameters are set as 
 $\tan \theta_{\odot}=2.6\times 10^{-1}$ and $\theta_{atm}=\pi/4$. 
Majorana phases $\phi_{1}$ and $\phi_2$ are normalized by $2\pi$.
}
\label{Type B2}
\end{figure}

Next, we see 
 the effect of $U_{e3}$ in Type B.
In general the change of $U_{e3}$ induces 
 the changes of $U_{e1}$ and $U_{e2}$, which 
 causes the deviation of $\langle m \rangle$ in Eq.(\ref{<m>}). 
Since 
 we only know the upper bound of $|U_{e3}|$ as,
 $|U_{e3}|<0.16$, 
 we compare 
 the cases of $|U_{e3}|=0$ in Fig.\ref{Type B2} (a),
 and $|U_{e3}|=0.1$ in Fig.\ref{Type B2} (b).
 The value of 
 $\langle m \rangle$ becomes small as 
 the magnitude of $U_{e3}$ becomes large. 
However, this effect is negligibly small
shown in Fig.\ref{Type B2} (a) and (b).  
So the effect of the magnitude of $U_{e3}$
 to $\langle m \rangle$
 is negligible.
The crucial point of this result 
 comes from $m_3\simeq 0$ in Eq.(\ref{<m>}).  
In the case of non-zero value of $U_{e3}$, 
 $CP$ phase $\delta$ is physical, but 
 this effect 
 is also negligible.

We also estimete the effect of 
 the deviation from $m_3=0$. 
Since  
 $m_{3}$ should be much smaller than the value of 
 $|m_1|\simeq 5.6\times 10^{-2}\;\mathrm{eV}$ in 
 Type B, we take $m_3=0.1\times m_1$.
We find 
that this effect is also negligible.

As shown above, 
 the values of 
 $\langle m \rangle$
 cannot be larger than 
 $0.06$ eV in any parameter sets in Type B.  
The region where $\langle m \rangle > 0.05$ eV 
 only exists just around B2. 
This magnitude of $\langle m \rangle$ is 
 just edge of the allowed region of 
 experimental value of $95\%\;\mathrm{c.l.}$ 
 in Eq.(\ref{result}), and 
 far from the best fit value, $0.4$ eV. 
The improvement of experiments 
 might dispose the possibility 
 of type B in the near future.

\subsection{Type C}

The neutrino masses are degenerate in Type C, and 
 we set the value of the degenerate mass as $m_\nu$.
In Ref.\cite{Altarelli}, 
 Type C mass hierarchy is classified to four subgroups, 
 C0, C1, C2 and C3, by relative signs $m_1$, $m_2$ and $m_3$.
\begin{eqnarray*}
(\tilde{m}_1,\tilde{m}_2,\tilde{m}_3)=
\left\{
\begin{array}{rrrccccc}
m_\nu(&1,&1,&1)&\quad&(\phi_1,\phi_2)=(0,0)&\quad&(\mathrm{Type\;C0})\\
m_\nu(&-1,&1,&1)&&(\phi_1,\phi_2)=(0,\pi)&&(\mathrm{Type\;C1})\\
m_\nu(&1,&-1,&1)&&(\phi_1,\phi_2)=(\pi,0)&&(\mathrm{Type\;C2})\\
m_\nu(&-1,&-1,&1)&&(\phi_1,\phi_2)=(\pi,\pi)&&(\mathrm{Type\;C3})
\end{array}
\right.
\end{eqnarray*}
\begin{figure}
\setlength{\unitlength}{1cm}
\begin{picture}(13,13)
\put(0,7.5){\includegraphics[scale=.7]{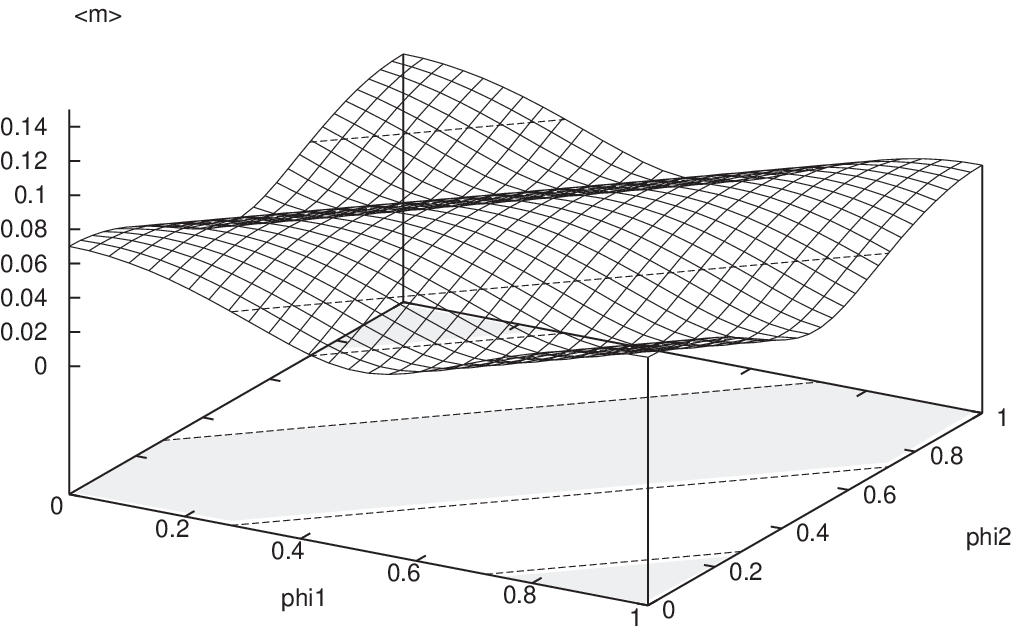}}
\put(4,6.5){(a)}
\put(8,7.5){\includegraphics[scale=.7]{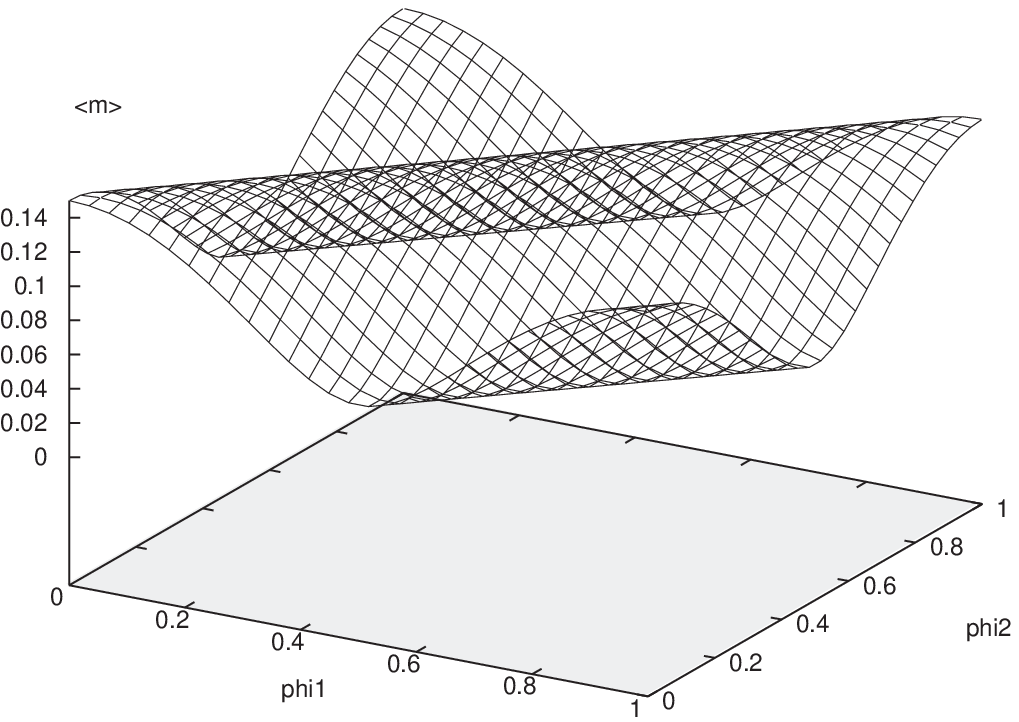}}
\put(12,6.5){(b)}
\put(0,1){\includegraphics[scale=.7]{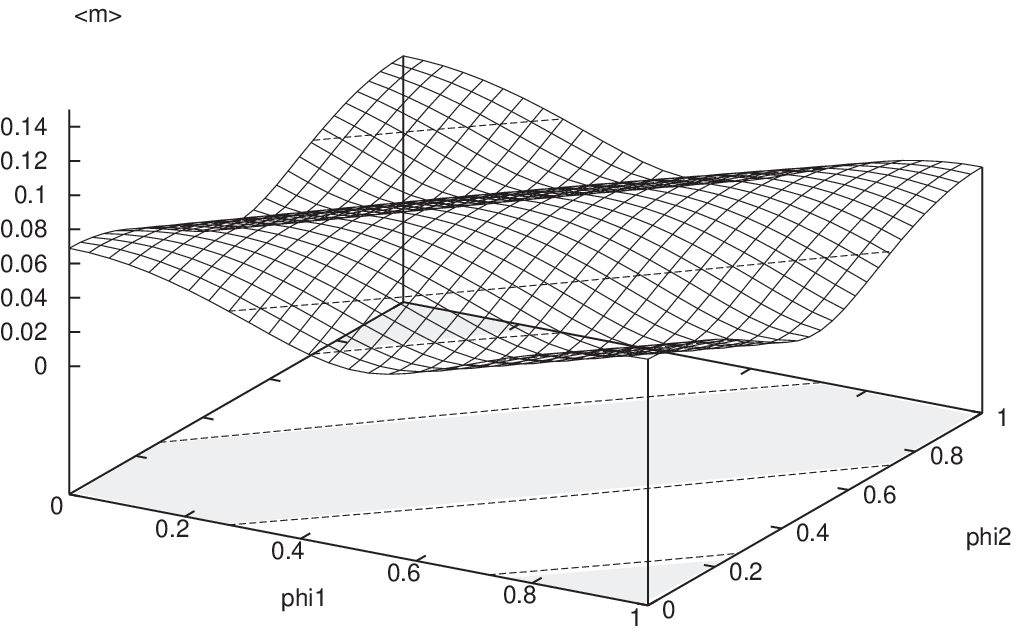}}
\put(4,0){(c)}
\end{picture}
\caption{The values of $\langle m \rangle$ in Type C.
Each graph shows 
(a): $m_\nu=0.07$ eV, $U_{e3}=0$, (b): $m_\nu=0.15$ eV, $U_{e3}=0$
and
(c): $m_\nu=0.07$ eV, $|U_{e3}|=0.1$, $\delta=\pi/2$.
Other parameters are set as 
 $\tan \theta_{\odot}=2.6\times 10^{-1}$ and $\theta_{atm}=\pi/4$. 
Majorana phases $\phi_{1,2}$ are normalized by $2\pi$.
}
\label{TypeC1}
\end{figure}
Seeing the zeroth order neutrino mass matrices 
 in Table \ref{table:mass}, 
 we suppose naively 
 only 
 Type C0 and C3 might explain $0\nu\beta\beta$
 decay experiments because the (1,1) element of 
 mass matrix is of $\mathcal{O}(m_\nu)$. 
Can Type C1 and C2 
 not really explain Eq.(\ref{result})?

In Fig.\ref{TypeC1} (a) and Fig.\ref{TypeC1} (b), 
 the values of $\langle m \rangle$ are shown 
 in the case of $m_{\nu}=0.07$ eV and
 $m_{\nu}=0.15$ eV, respectively.  
Where we take 
   $U_{e3}=0$.
It is worth noting that 
 the values of $\langle m \rangle$ strongly 
 depend on the value of $m_\nu$.
In the case of $m_\nu=0.15$ eV in Fig.\ref{TypeC1} (b), 
 any values of $\phi_{1,2}$ can 
 explain the $0\nu\beta\beta$ decay in Eq.(\ref{result}). 
The suitably large value of $\langle m \rangle$ 
 satisfying Eq.(\ref{result}) 
 can be obtained in Type C1 and C2, 
 although zeroth order forms of neutrino mass matrices 
 in Table \ref{table:mass} show 
 tiny $(1,1)$ elements in 
 Type C1 and C2. 
This is simply because
 that
 $\theta_{\odot}$ deviates from the maximal mixing tinily
 and that 
 all elements, of course including the $(1,1)$ component, 
 become large as $m_\nu$ becomes large. 
Thus the suitable large $\langle m \rangle$ 
 can be obtained even if the $(1,1)$ component 
 is of order zero 
 in the zeroth order approximation 
 in Table 2.


As for the effects of $|U_{e3}|$ and $CP$ phase $\delta$, 
 they should induce the sensitive effects to 
 the magnitude of $\langle m \rangle$ in Type C,
 because the value of $m_{3}$ is large comparing 
 to Type B.
We show the case of
$CP$ phase 
$\delta =\pi/2$ (Fig.\ref{TypeC1} (c)), 
where 
we take $|U_{e3}|=0.10$ and $m_\nu=0.07$ eV.
{}From Figs.\ref{TypeC1} (a) 
 and (c),
 we find that 
 the region around $(\phi_1,\phi_2)=(\pi,\pi)$ 
 receives the sensitive effects from $|U_{e3}|$ and $\delta$,  
 but they are not so large. 
We also estimate the effects in 
 the case of $\delta=0$,
 but we find that this effect is negligible.
 
As shown above,   
 the main contribution which lifts 
 the value of $\langle m \rangle$ 
 exists in $m_\nu$ in Type C. 
Thus we can conclude that 
 the suitably large $m_\nu$ makes 
 the value of $\langle m \rangle$ be large enough to 
 satisfy Eq.(\ref{result}) 
 in any values of Majorana phases 
 $\phi_1$ and $\phi_2$ in Type C.


\section{Conclusion}

A recent paper\cite{db} 
 announces the evidence of 
 $0\nu\beta\beta$ decay, and  
 the value of $\langle m \rangle$ 
 is large as Eq.(\ref{result}). 
In this paper, 
 we have estimated  
 the value of 
 $\langle m \rangle$ 
 according to 
 the neutrino mass hierarchy patterns, 
 Type A, B and C. 
We have searched which mass hierarchy is 
 consistent 
 with the recent $0\nu\beta\beta$ decay experiments. 
We have also analyzed the deviation from 
 the zeroth order forms of neutrino mass matrices 
 with the fluctuations of small physical parameters 
 in the base of the Majorana $CP$ phases. 

The results are the followings. 
Type A cannot 
 explain the $0\nu\beta\beta$
 results in Eq.(\ref{result}).
In Type B, 
 the region where $\langle m \rangle > 0.05$ eV 
 exists around B2 in the parameter space of 
 Majorana $CP$ 
 phases. 
In Type C, 
 the suitably large $m_\nu$ makes 
 the value of $\langle m \rangle$ be large enough to 
 satisfy Eq.(\ref{result}) 
 in any values of Majorana phases 
 $\phi_1$ and $\phi_2$, contrary to 
 the zeroth order estimations.  
The effects of $|U_{e3}|$ and $\delta$ are 
 not be significant in all types of neutrino mass
 hierarchies.

{\small
\begin{table}[b]
\begin{minipage}{8.3cm}
\hspace{-1.7cm}
\begin{tabular}{|c|c|c|}
\hline     & &Neutrino  \\
           & $\widetilde{M}_\nu$  &mass matrix        \\
\hline & & \\
 A &\small $diag.(0,0,1)$ &
$\left[ 
\begin{array}{ccc} 
 0&0&0\\
 0&\um&\um\\
 0&\um&\um
\end{array}
\right]$ 
 \\ & &  \\
\hline & &  \\
 B1 &\small$diag.(1,-1,0)$&
\small$\left[
\begin{array}{ccc}
 0&-\sq&\sq\\
 -\sq&0&0\\
 \sq&0&0
\end{array}
\right]$ 
 \\ & &  \\
\hline & &  \\ B2 &\small$diag.(1,1,0)$&
$\left[
\begin{array}{ccc}
 1&0&0\\
 0&\um&-\um\\
 0&-\um&\um
\end{array}
\right]$ 
 \\ & &  \\
\hline 
\end{tabular}
\vspace{1.7cm}
\end{minipage}
\begin{minipage}{8cm}
\hspace{-.7cm}
\begin{tabular}{|c|c|c|}\hline
&&\\
C0 &\small $diag.(1,1,1)$&
$\left[
\begin{array}{ccc}
 1&0&0\\
 0&1&0\\
 0&0&1
\end{array}
\right]$ 
 \\ & &  \\
\hline & & \\ C1 &\small $diag.(-1,1,1)$&
\small$\left[
\begin{array}{ccc}
 0&\sq&-\sq\\
 \sq&\um&\um\\
 -\sq&\um&-\um
\end{array}
\right]$ 
 \\ & & \\
\hline & & \\ C2 &\small$diag.(1,-1,1)$&
\small$\left[
\begin{array}{ccc}
 0&-\sq&\sq\\
 -\sq&\um&\um\\
 \sq&\um&\um
\end{array}
\right]$ 
 \\ & & \\
\hline & & \\ C3 &\small $diag.(-1,-1,1)$&
$\left[
\begin{array}{ccc}
-1&0&0\\
 0&0&1\\
 0&1&0
\end{array}
\right]$ 
 \\ & & \\
\hline
\end{tabular}
\end{minipage}
\caption
{The zeroth order neutrino mass matrices. 
In Type A and B, 
the eigenvalues of $\widetilde{M}_\nu$ and the neutrino mass matricies 
are normalized by $\sqrt{\Delta m_{atm}^2}.$
In Type C,
they are normalized by $m_\nu$.
}
\label{table:mass}
\end{table}
}

\section*{Acknowledgment}
We would like to thank M. Tanimoto for letting authors 
 know the papaer of neutrinoless double $\beta$. 
This work is supported in part by the Grant-in-Aid for Science Research,
 Ministry of Education, Science and Culture, Japan (No. 12740146 ).



\begin{thebibliography}{99}

\bibitem{Kamiokande}
Y.~Fukuda {\it et al.}  [Kamiokande Collaboration],
Phys.\ Rev.\ Lett.\ {\bf 77} (1996) 1683.

\bibitem {SKatm}
Y.~Fukuda {\it et al.}  [Kamiokande Collaboration],
Phys.\ Lett.\ B {\bf 335} (1994) 237;
\\
Y.~Fukuda {\it et al.}  [Super-Kamiokande Collaboration],
Phys.\ Rev.\ Lett.\ {\bf 81} (1998) 1562;
\\
T.~Kajita  [Super-Kamiokande Collaboration], 
in {\it Neutrino Physics and Astrophysics},
Proceedings of the XVIIIth International Conference on Neutrino
Physics and Astrophysics (Neutrino '98), June 4-9, 1998, Takayama,
Japan, edited by Y. Suzuki and Y. Totsuka,
(Elsevier Science B.V., Amsterdam, 1999) page 123;
Nucl.\ Phys.\ Proc.\ Suppl.\ {\bf 77}, 123 (1999).





\bibitem{SKsolar}
Y.~Fukuda {\it et al.}  [Super-Kamiokande Collaboration],
\\
Phys.\ Rev.\ Lett.\ {\bf 81} (1998) 1158;
Erratum-ibid.\ {\bf 81} (1998) 4279;
\\
Phys.\ Rev.\ Lett.\ {\bf 82} (1999) 2430;
\\
Phys.\ Rev.\ Lett.\ {\bf 82} (1999) 1810.

\bibitem{SNO}

Q.~R.~Ahmad {\it et al.}  [SNO Collaboration],
Phys.\ Rev.\ Lett.\  {\bf 87}, 071301 (2001).




\bibitem{Cl-Homestake}
K.~Lande {\it et al.},
Astrophys.\ J.\  {\bf 496} (1998) 505.


\bibitem{Ga-Gallex-and-GNO}
V.~N.~Gavrin  [SAGE Collaboration],
Nucl.\ Phys.\ Proc.\ Suppl.\  {\bf 91} (2001) 36;
\\
E.~Bellotti,
Nucl.\ Phys.\ Proc.\ Suppl.\  {\bf 91} (2001) 44.


\bibitem{post-SNO-analysis}
V.~D.~Barger, D.~Marfatia and K.~Whisnant,
Phys.\ Rev.\ Lett.\  {\bf 88}, 011302 (2002);
\\
G.~L.~Fogli, E.~Lisi, D.~Montanino and A.~Palazzo,
Phys.\ Rev.\ D {\bf 64}, 093007 (2001);
\\
J.~N.~Bahcall, M.~C.~Gonzalez-Garcia and C.~Pena-Garay,
JHEP {\bf 0108}, 014 (2001);
\\
A.~Bandyopadhyay, S.~Choubey, S.~Goswami and K.~Kar,
Phys.\ Lett.\ B {\bf 519}, 83 (2001).

\bibitem{db}
H.~V.~Klapdor-Kleingrothaus, A.~Dietz, H.~L.~Harney and I.~V.~Krivosheina,
Mod.\ Phys.\ Lett.\ A {\bf 16}, 2409 (2002)

\bibitem{dark}
V.~Barger, S.~L.~Glashow, D.~Marfatia and K.~Whisnant,
arXiv:hep-ph/0201262.






\bibitem{seesaw}
T.~Yanagida,
{\it ``Horizontal Symmetry And Masses Of Neutrinos''},
Prog.\ Theor.\ Phys.\  {\bf 64} (1980) 1103,
and in Proceedings of the 
{\it ``Workshop on the Unified Theory and the Baryon Number in the
  Universe''}, Tsukuba, Japan, Feb 13-14, 1979, 
  Eds. O.~Sawada and A.~Sugamoto, KEK report KEK-79-18, p. 95;
\\
M.~Gell-Mann, P.~Ramond and R.~Slansky,
{\it in} ``Supergravity''
(North-Holland, Amsterdam, 1979)
{\it eds.} D.Z. Freedman and P. van Nieuwenhuizen,
Print-80-0576 (CERN).


%
Phys.\ Lett.\ B {\bf 161} (1985) 141.

\bibitem{Altarelli}
G.~Altarelli and F.~Feruglio,
Phys.\ Rept.\  {\bf 320}, 295 (1999).
\bibitem {MNS}Z. Maki, M. Nakagawa, and S. Sakata, Prog. Theor. Phys.
	 {\bf 28} (1962) 870.

\bibitem{CHOOZ}
M.~Apollonio {\it et al.}  [CHOOZ Collaboration],
Phys.\ Lett.\ B {\bf 466}, 415 (1999)




\end{thebibliography}
\end{document}